# Radiative Reactions and Coherence Modeling in the High-Altitude Electromagnetic Pulse

**Charles N. Vittitoe**
*Sandia National Laboratories, P.O. Box 5800, Albuquerque, New Mexico 87185-5800*

**Mario Rabinowitz**
*Electric Power Research Institute, Palo Alto, California 94303*
Inquiries to: Mario Rabinowitz
*Armor Research, 715 Lakemead Way, Redwood City, CA 94062*
Mario715@earthlink.net

**Abstract**

A high-altitude nuclear electromagnetic pulse (EMP) with a peak field intensity of $5 \times 10^4$ V/m carries momentum that results in a retarding force on the average Compton electron (radiating coherently to produce the waveform) with magnitude near that of the geomagnetic force responsible for the coherent radiation. The retarding force results from a self-field effect. The Compton electron interaction with the self-generated magnetic field due to the other electrons accounts for the momentum density in the propagating wave; interaction with the self-generated electric field accounts for the energy-flux density in the propagating wave. Coherent addition of radiation is also quantitatively modeled.

## INTRODUCTION

Electrons propagating in air can radiate an associated electromagnetic pulse (EMP).[1] Over two decades ago, Conrad Longmire [2] at Los Alamos, and William Karzas and Richard Latter at RAND Corporation [3] developed a model for the EMP generated by a high-altitude nuclear explosion. The basic mechanism for this pulse generation is coherent radiation from the Compton electrons gyrating about the Earth's geomagnetic field. The pulse is produced in the gamma-ray absorption region of the atmosphere at an altitude of 20 to 40 km and radiates to large distances within line of sight of the gamma-ray source. In 1975 Sherman et al., Bell Laboratories, represented the radiating electric field as a double exponential pulse with a peak intensity of 50 kV/m, a rise time near 10 ns, and a fall to one-half the peak value requiring over 100 ns. [4] We examine the forces acting on the Compton electrons and find that the peak field results in a significant retarding force. We inquire if this retarding force is a significant, neglected force that reduces expected peak field values because it is not obvious that this force is accounted for when current densities are modeled. Further motivation results because the relativistic form of the radiation reaction force is not readily calculable and because this reaction force has not been included in the models. A significant reduction in peak field strength below 50 kV/m would reduce the EMP threat and would save considerable taxpayer dollars spent hardening systems to the EMP environment.

## LARGE RETARDING FORCE RESULTING FROM MOMENTUM IN THE PROPAGATING WAVE

Consider an EMP source region that is ~ 10 km in depth along a line of sight vertically downward at the equator. The coherent addition of radiation propagating downward along the line of sight results in energy and momentum being taken from the Compton electrons. A much weaker signal is radiated backward along the line of sight, and so a net force in the upward direction acts upon the ensemble of coherent electrons. The force is taken as the net time rate of change in the momentum carried by the radiation.

Let the downward propagating wave have a peak electric field of 50 kV/m. The associated energy flux density is given by the Poynting vector $|\vec{E} \times \vec{H}|$ = 6.6 x $10^6$ W/m$^2$. At the peak energy flux density, the corresponding momentum flux density (in a plane-wave approximation) is given by (6.6 x $10^6$ W/m$^2$)/c = 2.2 x $10^{-2}$ N/m$^2$, or approximately 3 x $10^{-6}$ psi exerted upon a perfectly absorbing screen. This momentum flux density is equal and opposite to the rate of change in momentum per unit area of the ensemble of electrons producing the radiation. Further division by c gives the momentum density at the peak of the radiating wave g = 7.3 x 10-11 Ns/m³.

If there are $n_c$ coherent, radiating electrons per square meter within the source region, then the peak radiation-associated force on the average electron is
$$F_{rad} = 2.2 \times 10^{-2} / n_c \quad N.$$
Longmire [5] has estimated that about 5 x 10¹⁸ Compton electrons contribute coherently, spread over an area A ~ π(400 m)², giving a rough estimate of $n_c$ ~ 10¹³ e/m². This implies that the radiation--associated retarding force on the average electron is approximately equal to 2.2 x 10⁻¹⁵ N at the peak in the radiated power. Because of the 1/r reduction in the radiated field, the force in the source region will be larger by a few tens of percent. This force will now be compared with the other forces on the Compton electrons to determine if it is negligible.

Solving the Compton electron equations of motion is a critical step in determining the basic drivers for Maxwell's equations and is necessary for quantitative estimates of the radiating fields. Forces acting upon these Compton electrons appear in the equations of motion. The Lorentz force gives contributions associated with the E and B fields. With the air density represented by ρ, and the sea-level value by $ρ_o$, at cw frequency less than or equal to 100 MHz and at altitudes less than 60 km the electric fields are limited by air breakdown to values of E < 3 x 10⁶ ρ/$ρ_o$ V/m. At an altitude of 30 km, near the center of the source region for high-altitude EMP, this force is $F_E$ = eE < 6.6 X 10⁻¹⁵ N. The Earth's geomagnetic field is approximately equal- to 6 X 10⁻⁵ T, which implies that magnetic forces can reach $F_{Bgeo} = e|\vec{v} \times \vec{B}| \leq 2.9 \times 10^{-15}$ N. With W representing the electron kinetic energy, the effective force produced by stoppingpower effects (at β=v/c=0.9) is represented as [6] $F_{stop}$ = dW/dx=(2 MeV cm²/g)ρ < 5.4 x 10⁻¹⁶ N, where the 30-km air density is used for the estimate. The problem is time dependent. The initial electric field is zero, velocities are changing, and air densities vary. Hence, each of these forces should be considered in detail in the Compton electron equations of motion, The peak radiation-associated retarding force of approximately $2 \times 10^{-15}$ N is significant; it must be taken into account when the electron equations of motion are solved. Rabinowitz has presented independent arguments that show that this retarding force is comparable to the geomagnetic force that produces the coherence and the radiation.[7]

The momentum density imparted to the radiating electrons can be reduced somewhat by including the radiation that propagates upward. The downward-directed pulse has a width $\Delta t_d$. The upward-directed pulse is spread over a much longer time, dependent upon a coherence length $l$. As a rough estimate, the width of the upward directed pulse is $\Delta t_u = \Delta t_d + 2l/c$. Since $\Delta t_d \ll 2l/c$, $\Delta t_u / \Delta t_d \sim 2l/(c \Delta t_d)$. The net impulse is

$$F_d \Delta t_d - F_u \Delta t_u = F_d \Delta t_d \left( 1 - \frac{2 \ell F_u}{c \Delta t_d F_d} \right).$$

This impulse is directed upward on the radiating electrons and furnishes the downward-directed momentum density in the propagating wave. So the effective retarding force is reduced by the factor within the brackets. With a coherence length $l \sim 10^4$ m, $\Delta t_d \sim 10^{-8}$ sec, and with an estimated $F_u = 10^{-6} F_d$ corresponding to an upper bound on the upward propagating electric field approximately equal to $10^{-3}$ times the coherent downward signal, the momentum density given to the electrons is reduced by 0.7%. The angular distribution of radiation from an isolated 1-MeV electron in helical motion canalso be used to estimate the $F_u / F_d$ ratio. The ratio of the angular distributions of radiated power evaluated in the backward direction to that in the forward direction (relative to the velocity) is then approximately $10^{-5}$. Coherence effects acting constructively in the downward direction drastically lower this estimate. The net momentum transfer is large. The backward wave does not significantly reduce the estimate of the radiation-associated force.

Consider properties of the force to see whether the force is already included, perhaps as part of the self-field effects. The reaction force is directed vertically upward and should be dependent upon the horizontal accelerations responsible for the coherent radiation (for our vertically downward line of sight). The self-fields include horizontal electric fields, but can result in a vertical force only through substitution of the associated accelerations into radiation reaction types of force where components appear antiparallel to the particle velocity. Since the electron velocity is nearly vertically downward the v x B (where B is the geomagnetic field or the self-generated magnetic field in the propagating wave) gives only a small vertical force, resulting from the horizontal component of v. In addition, the cross product does not change the particle energy, and so it cannot account for the energy carried away by the radiation. The $F_{stop}$ is a statistical, incoherent effective force; the energy deposited forms ionization and excitation, not radiation. It is not obvious that $F_{rad}$ can be accounted for by self-field effects or by energy-loss effects already included within the equations of motion. Since radiation reaction forces on individual electrons are not included in the equations of motion, we next examine if this accounts for the large retarding force.

## RELATIVISTIC FORM AND NUMERICAL ESTIMATES FOR THE RADIATION REACTION FORCE

The radiation reaction force has traditionally been neglected in EMP applications. This force results because the radiated wave carries energy, momentum, and angular momentum. The nonrelativistic form for the force is (in SI units where the subscript denotes radiation reaction)[8] $F'_{rr} = e^2 / 6\pi\varepsilon_0 c^3 \, da/dt$. Because the Compton electron speeds are ~0.9c, relativistic effects must be included. Here we examine several forms for the relativistic radiation reaction force. Some are judged nonphysical. We concentrate on two forms: the first is associated with the generation of electromagnetic four-vector momentum $\left(F^\mu_{rr}\right)$; the second is the Schott force on an individual radiating electron $(S^\mu)$.

One covariant expression for the radiation reaction force is [8]

$$F_\mu^{rad} = \frac{e^2}{6\pi\varepsilon_0 c^3 m}\left[\frac{d^2 p_\mu}{d\tau^2} + \frac{p_\mu}{m^2 c^2}\left(\frac{dp_\nu}{d\tau}\frac{dp^\nu}{d\tau}\right)\right], \quad (1)$$

where $\tau$ is the particle's proper time. This is consistent with the form given by Rohrlich [9] for the "radiation reaction" force,

$$\Gamma^\mu = \frac{e^2}{6\pi\varepsilon_0 c^3}\left(\ddot{a}^\mu - c^{-2} a^\nu a_\nu v^\mu\right), \quad (2)$$

The "dot" over a variable denotes a total derivative with respect to $\tau$. The sign change results from a sign change in the definition of the metric tensor (that enters in the $a^\nu a_\nu$ term). The Rohrlich form has greater clarity in showing that the velocity term in the reaction force is directed opposite to the particle velocity. A general four-vector $f^\nu$ is spacelike (or timelike) when $f^\nu f_\nu$ is positive (or negative). The $a^\nu$ is always spacelike. Cohn [10] presents a clear heuristic derivation of $\Gamma^\mu$.

These forms for a radiation reaction force ($\Gamma^\mu$ and $F_\mu^{rad}$) do not readily lend themselves to a physical interpretation. The Lorentz invariant energy radiation rate is (using Rohrlich's metric tensor)[9]

$$P = \frac{e^2}{6\pi\varepsilon_0 c^3} a^\nu a_\nu = \frac{e^2 \gamma^4}{6\pi\varepsilon_0 c^3}\left[a^2 + \frac{\gamma^2}{c^2}(\vec{v}\bullet\vec{a})^2\right]. \quad (3)$$

If $a^\nu = 0$ but $\dot{a}^\mu \neq 0$, then $P = 0$ with Eq. (2) showing that $\Gamma^\mu \neq 0$ for an instant. However, the radiation reaction force should be zero whenever the radiated power (P) is 0. When $c^2 \dot{a}^\mu = a^\nu a_\nu v^\mu$ with $a^\nu \neq 0$, then Eq. (2) gives $\Gamma^\mu = 0$; yet $P > 0$. Thus the label radiation reaction force is a misnomer for $\Gamma^\mu$. Later we show that the scalar product of our forms for the radiation reaction force and the velocity four-vectors is the radiated power P. Because $a^\mu v_\mu = 0$, $\dot{a}^\mu v_\mu = -a_\mu a^\mu$, and $v^\mu v_\mu = -c^2$, we have $\Gamma^\mu v_\mu = 0$, illustrating the orthogonality that is also required of the total force.

The radiation reaction force has also been identified by saying that this force is equal and opposite to the rate at which electromagnetic four-vector momentum is emitted.[9]

$$F_{rr}^\mu = -\frac{1}{c^2} P v^\mu, \qquad (4)$$

where P is the Lorentz invariant energy radiation rate as defined by Eq. (3). This gives the force a direct physical interpretation. Note that it is identical to the second term in the definition of $\Gamma^\mu$ in Eq. (2) and is the force we will retain as the radiation reaction force. The rate at which the force does work is

$$F_{rr}^\mu v_\mu = P. \qquad (5)$$

Equation (3) is used to estimate P for an individual radiating electron. Let the Lorentz factor of special relativity, $\gamma = 3$ (kinetic energy=1 MeV, $\beta = 0.94$) for a Compton electron in instantaneous circular motion in a magnetic field $B = 60$ µT. Then $v \cdot a = 0$, $a = eBv/\gamma m = 9.92 \times 10^{14}$ m/s$^2$, and the instantaneous power radiated is $P = e^2 \gamma^4 a^2 / 6\pi\varepsilon_o c^3 = 4.5 \times 10^{-22}$ W. Equation (4) shows that the spatial portion of $F_{rr}^\mu$ associated with this P is much less than the $F_E$, $F_{Bgeo}$, and $F_{Stop}$ estimated earlier and is negligible in the electron equations of motion. If this power were radiated incoherently by $n_c = 10^{13}$ e/m$^2$, the total power per unit area would be $n_c P = 4.5 \times 10^{-9}$ W/m$^2$, much less than the $6.6 \times 10^6$ W/m$^2$ in the EMP peak. If this power were radiated coherently by each of (the earlier noted) approximately $5 \times 10^{18}$ contributing electrons, the power becomes much greater than that in the EMP peak. However the contributing electrons are only partially in phase. The coherent addition that gives $6.6 \times 10^6$ W/m$^2$ at the EMP peak led us to the $2.2 \times 10^{-15}$ N force on the average electron that we found earlier.

The estimate from Eq. (4) showed that the individual Compton electrons generate a negligible $F_{rr}^\mu$. We now review some coherence effects to investigate their consequences on these forces. The electric field generated by one electron is given by (Ref. 8 with conversion to SI units)

$$E(x,t) = \frac{e}{4\pi\varepsilon_o} \left[ \left( \frac{n - \beta}{\gamma^2 (1 - \beta \cdot n)^3 R^2} \right)_{ret} + \left( \frac{n \times [(n - \beta) \times \dot\beta]}{(1 - \beta \cdot n)^3 Rc} \right)_{ret} \right].$$

Here, $\beta = v/c$, $\vec{R}$ is the vector from the source electron to the point of observation, x is the observation point, n is a unit vector along R, and $R = |\vec{R}|$. The subscript, ret, denotes evaluation at retarded time. To estimate this electric field in the radiation zone, where the 1/R term dominates, we again take our γ = 3 electron in instantaneous circular motion in B=60 μT. We examine the transverse E radiating in the forward direction. Here n is parallel to β and $\dot{\beta} = eBv/\gamma mc$. The distance R is taken as 35 km (roughly the distance from the gamma-ray absorption region to the earth directly under our burst). Our sample electron generates a peak electric field radiated in the forward direction, E=1.3 x $10^{-13}$ V/m. If a group of electrons is radiating coherently in this forward direction, then $N_c$=4 x $10^{17}$ electrons give E = 5 x $10^4$ V/m, the peak EMP. At a lower average electron energy, because of the smaller β more electrons would be required to furnish the same field. A lower estimate for $N_c$ can be obtained by reducing R to perhaps 30 km. The power and the electric field radiated by the ensemble is increased by the coherence. The associated radiation forces must also increase.

We broaden our search for the large radiation reaction force by following Rohrlich [9] and introducing the Schott force. Rohrlich defines a force four-vector as

$$K^\mu(\tau) = F^\mu_{in} + F^\mu_{ext} - \frac{1}{c^2} P v^\mu, \qquad (6)$$

that is identified as the *effective force response for the acceleration of the particle*. The terms are the following: the force resulting from incident radiation, the external forces, and the radiation reaction force previously encountered in Eq. (4) (which is the negative of the rate at which electromagnetic four-momentum is emitted). The Lorentz-Dirac equation of motion [9]

$$ma^\mu = F^\mu_{in} + F^\mu_{ext} + \Gamma^\mu \qquad (7)$$

can now be written in the form

$$K^\mu(\tau) = ma^\mu - \frac{e^2}{6\pi\varepsilon_o c^3} \dot{a}^\mu. \qquad (8)$$

The $F^\mu_{rr}$ portion of $\Gamma^\mu$ in Eq. (7) contributes (along with $F^\mu_{in}$ and $F^\mu_{ext}$) to the $ma^\mu$ term; it does not appear explicitly in Eq. (8). Rather than a strict radiation reaction force ($F^\mu_{rr}$), we have identified the Schott term $e^2 \dot{a}^\mu / (6\pi\varepsilon_o c^3)$ by which the effiective force differs from $ma^\mu$ (see Ref. 9, p. 146 ff):

$$S^\mu = \frac{e^2}{6\pi\varepsilon_o c^3} \dot{a}^\mu, \tag{9}$$

where $-S^\mu v_\mu = K^\mu v_\mu = F_{rr}^\mu v_\mu = P$. In addition to $\Gamma^\mu$ and $F_{rr}^\mu$ the Schott term in certain cases may also be called a radiation reaction force. $S^\mu$ is the first term of $\Gamma^\mu$ in Eq. (2). For the physical reasons stated above, we prefer not to call $\Gamma^\mu$ a radiation reaction force. This label is retained for $F_{rr}^\mu$ even though it is not orthogonal to the velocity four-vector (since it is not the total force). $S^\mu$ has nonphysical aspects in common with $\Gamma^\mu$ as well as being being nonorthogonal to the velocity four-vector. A negative $S^\mu$ indicates a four-vector that opposes the $ma^\mu$. Its neglect gives an overestimated instantaneous acceleration in Eq. (8).

Rohrlich [9] gives the $\dot{a}^\mu$ in three-vector form in terms of laboratory time, where $dt/d\tau = \gamma$;

$$\dot{a}^\mu(t) = \left[ \dot{a}^0; \gamma^3 \frac{da}{dt} + 3\gamma^5 (\vec{v} \cdot \vec{a}) \frac{a}{c^2} + \dot{a}^0 \frac{v}{c} \right], \tag{10}$$

where

$$\dot{a}^0(t) = \frac{\gamma^5}{c} \left[ v \cdot \frac{da}{dt} + a^2 \right] + \frac{4\gamma^7}{c^3} (\vec{v} \cdot \vec{a})^2. \tag{11}$$

The spatial three-vector form for the Schott term is

$$S(t) = + \frac{e^2}{6\pi\varepsilon_o c^3} \left[ \gamma^3 \frac{da}{dt} + 3\gamma^5 (\vec{v} \cdot \vec{a}) \frac{a}{c^2} + \dot{a}^0 \frac{v}{c} \right]. \tag{12}$$

We can now estimate the Schott force associated with several characteristics of the Compton electron motion. One of the large accelerations that might be experienced is produced, by the radial electric field. Using the breakdown field estimated earlier, we find that the acceleration is approximately equal to $F_E / (\gamma^3 m) \leq 6.6 \times 10^{-15} / 9.1 \times 10^{-31}$ = $7.25 \times 10^{15}$ m/s$^2$ = $a_1$ when $\gamma = 1$. The subscript on $a$ indicates the value of $\gamma$ used for evaluating the acceleration $a_1$. The $\gamma^3$ factor occurs because this acceleration is longitudinal, that is, antiparallel (or parallel) to the velocity. If the breakdown field were transverse, the acceleration would be $F_E / (\gamma m) = a_1 / \gamma$. With this latter acceleration we examine upper limits for the three terms in the Schott force.

For the transverse acceleration with $\gamma = 3$, $\Delta t = 10^{-8}$ s, and with the inconsistent but bounding assumption that v is antiparallel to a, the terms are bounded by

$$\gamma^3 da / dt \leq \gamma^2 a_1 / \Delta t = 6.52 \times 10^{24}, \tag{13}$$

$$3\gamma^5 (\beta \cdot a)a / c \leq 3\gamma^3 \beta a_1^2 / c = 1.33 \times 10^{25} \tag{14}$$

$$\beta \dot{a}^0 \leq \gamma^3 \beta a_1 (\gamma \beta / \Delta t + a_1 / c + 4\gamma^2 \beta a_1 / c) = 1.98 \times 10^{26}. \tag{15}$$

If the vectors are such that these terms add, then the upper bound is

$$|S(t)| \leq \frac{e^2}{6\pi\varepsilon_0 c^3} \left( 2.18 \times 10^{26} \text{ m/s}^3 \right) = 1.24 \times 10^{-25} \text{ N}. \tag{16}$$

These contributions to the Schott force are several orders of magnitude below the $F_E$, $F_{Bgeo}$, and $F_{Stop}$ estimated earlier, and are negligible. For GeV rather than MeV electron energies, the Schott force becomes comparable to the other forces even for just the contributions examined above. At larger energies the Schott force and the $F_{rr}$ will be the dominant forces.

The geomagnetic force results in an acceleration that contributes to the radiation reaction force. In this case $\beta \cdot a = 0$, reducing the upper bounds found earlier. The resulting acceleration is smaller than the earlier estimate of radial acceleration; hence this contribution to the Schott force is also negligible. For particle motion in a circular orbit, we later find that $\dot{a}^0$ is zero, so the timelike component of $S^\mu$ is zero. For this orbit the spacelike component of $F_{rr}^\mu$ is also larger than the spacelike component of $S^\mu$. The ratio is $F_{rr} / S = \gamma^2 - 1$.

The arguments here have shown that the examined contributions to the Schott force (that result from the Compton electron motion in breakdown electric fields and in the geomagnetic field) are negligible. When a significant radiation (associated with other accelerations) must be accounted for as in the EMP case, then the associated

Schott terms might be large, and the resulting force could then dominate [7] the terms contributing to the Schott force and be comparable to the $F_E$, $F_{Bgeo}$, and $F_{Stop}$.

Let us consider ways in which coherence effects might increase the Schott force. When applied to a single electron traveling in a path like that experienced in the EMP case, the Schott force on an individual electron is expected to be approximately the same in magnitude as the tangential force necessary to keep a radiating particle traveling in a circle of radius r and radiating independently of other electrons. Here $\vec{v} \cdot \vec{a} = 0$, $da/dt = -a^2 v/v^2$, $a = v^2/r$, and $\dot{a}^0 = 0$. The Schott force becomes

$$S_\oplus(t) = \frac{e^2 \gamma (\gamma^2 - 1)}{6\pi\varepsilon_0 r^2 c} v, \tag{17}$$

and is the negative of the required tangential force. In terms of proper time $\tau$ (where $d/d\tau = \gamma d/dt$, $\gamma S(t) = S(\tau)$, the Schott force is in agreement with Rohrlich, [9] who has solved this problem with a different approach. Without the tangential force, the radiation rate is initially the same as for the actual orbit. The radiation loss then results in a reduction in v (and a reduction in r if the circular motion is provided by an imposed uniform magnetic field). Coherence effects can increase this force. If there is an ensemble of $N_c$ electrons radiating coherently as one body, then the total force on the ensemble will scale as the square of the charge, $(N_c e)^2$, and the force on the average electron is

$$F_{rad} = -N_c \frac{e^2 \gamma (\gamma^2 - 1)}{6\pi\varepsilon_0 r^2 c} v. \tag{18}$$

For a typical high-altitude EMP with v = 0.9c, $\gamma$ = 3, r=80 m, and $N_c > 10^{15}$ electrons, $F_{rad} > 10^{-15}$N. This is comparable to the geomagnetic force on each electron radiating to produce the EMP. Such a force cannot be neglected. However, Eq. (18) can only be applied approximately in the equations of motion. The Compton electrons are generated over a large region of space and only a subset within a quarter wavelength can radiate coherently as a single body. At f > 1 MHz, for example, $\lambda/4 < 75$ m; this is a small portion of the source region generating the high-altitude EMP. Even if the equation were valid for high-altitude EMP, a prescription would be needed to determine the $N_c$ appropriate for electrons at different positions in the gamma-ray absorption region as a function of time.

As an alternate view of the effects of coherence, consider Rohrlich's treatment [9] of a closed system of N charged particles. There, the Lorentz invariant form of the equation of motion for the kth particle is given by the Lorentz-Dirac equations for each of the particles (with proper asymptotic conditions as discussed by Rohrlich):

$$m_k a_k^\mu = F_{k,in}^\mu + \Gamma_k^\mu + F_{k,ret}^\mu \tag{19}$$

The acceleration of the kth particle is caused by three forces: $F_{k,in}$ produced by the electromagnetic fields incident upon the ensemble, the "radiation reaction" force $\Gamma^\mu$, and the force produced by all the other particles $F_{k,ret}^\mu$. The subscript ret indicates that the force from other particles must be evaluated in retarded time.

The $F_{k,ret}^\mu$ results from a sum over all the other particles, with the retarded time varying with each particle in the sum. As Rohrlich states, this term represents the only mutual interaction between the charged particles. The transverse electric field effects are built into this term, including whatever coherence is present in the local field. So a factor proportional to N does occur in the self-field term (with a constraint imposed by air breakdown). In this approach the radiation reaction force is not directly enhanced by a coherence effect. No multiplicative factor N appears in any part of the $\Gamma_k^\mu$ term in Eq. (19). Physically, the radiation reaction force on one electron cannot depend upon coherent addition of radiation if the addition occurs at a later time and at a location deeper in the atmosphere. When each electron radiates, it experiences a local electromagnetic field that is enhanced by coherence effects. This gives a correspondingly larger radiation reaction force.

In this section we have identified the radiation reaction forces $F_{rr}^\mu$ and $S^\mu$, and have chosen not to interpret $\Gamma^\mu$ as a radiation reaction force. The $F_{rr}^\mu$ on an individual electron was found to be negligible. The effects of breakdown electric fields and geomagnetic fields indicated a negligible $S^\mu$, even with expected coherence effects. We examined ways coherence effects enhance the radiation reaction forces. Earlier we showed that the high-altitude EMP peak amplitude implies a large radiation reaction. The following section shows that the large reaction is accounted for by a self-field effect.

## SELF-FIELD EFFECT

To account for the momentum in the propagating wave, we first review how the energy is generated by the transverse component of the current density. The argument is then extended to account for the momentum. The rate at which work is done per unit volume is given by $\vec{J} \cdot \vec{E}$. The particle kinetic energy is generating the field energy; $\vec{J} \cdot \vec{E}$ represents the power density for converting that kinetic energy into electromagnetic energy and thermal energy. The generation of thermal energy is accounted for by the sum of the stopping-power effects examined earlier and the conduction-current portion of $\vec{J} \cdot \vec{E}$. The radiating fields are intimately involved with the Compton electron energy. In 1974, Longmire showed (and later published in Ref. 11) that when the outgoing wave equation for the radiating field is solved self-consistently, the equation can lead to energy conservation. The energy thus given to the electromagnetic fields by the Compton electrons is shared by Joule heating in the conduction current and by the energy stored within the electromagnetic fields.

To illustrate explicitly the energy conservation, we consider a one-dimensional example that has some features similar to generation of high-altitude EMP. In Appendix A we derive the equation for the outgoing wave traveling in the direction $+\hat{x}$ (the circumflex indicates a unit vector in the associated direction). There we find

$$E_t(x, \tau) = -(Z_o / 2) \int_0^x J_t(x', \tau) dx' \tag{20}$$

for the transverse fields in free space at a given retarded time $\tau = t - x/c$. The notation is much the same as Longmire's [11] except that the current density $J_t$ includes the conduction as well as the convection terms provided by the Compton electrons.

Take the situation at a given $\tau$, where the transverse current density is given by

$$J_t(x,\tau) = \begin{cases} 0, & x < 0 \\ J_0, & 0 \leq x \leq L \\ 0, & x > L \end{cases} \tag{21}$$

A plane wave is being modeled, traveling in the x direction. The energy flux density in the plane wave (W/m 2) is given by

$$\frac{E_t^2}{Z_0} = \frac{Z_0}{4}\left[\int_0^x J_t(x',\tau)dx'\right]^2$$

$$\frac{E_t^2}{Z_0} = \frac{Z_0 J_0^2}{4}\begin{cases} 0, & x < 0 \\ x^2, & 0 \leq x \leq L \\ L^2, & x \geq L \end{cases} \tag{22}$$

The contribution made by the interval $(x_i, x_f)$ within (0, L) is $Z_0 J_0^2 (x_f^2 - x_i^2)/4$.

Each interval of constant width $\Delta x = x_f - x_i$ does not contribute equally to the transverse electric field intensity. To quantify this difference, we divide (0, L) into $l$ cells of length $\Delta x = L/\ell$, and treat the integral as a sum over the $l$ intervals. At the end of the nth cell

$$\int_0^{x_n} J_t(x',\tau)dx' = J_0 x_n = J_0 n \Delta x \tag{23}$$

and $E_t(x_n, \tau) = -Z_0 J_0 x_n / 2$. The corresponding energy flux density is

$$\frac{E_t^2}{Z_0} = \frac{Z_0 J_0^2}{4}\begin{cases} 0, & n \leq 0 \\ n^2(\Delta x)^2, & 1 \leq n \leq \ell \\ L^2, & x \geq \ell \end{cases} \tag{24}$$

The contribution of the nth cell to the energy flux density is given by

$$\left(\frac{E_t^2}{Z_0}\right)_n = \frac{Z_0 J_0^2}{4}(\Delta x)^2 \left[n^2 - (n-1)^2\right] \tag{25}$$

with the consistency check $\sum_{n=1}^{\ell}(2n-1) = \ell^2$. Coherence causes the nth cell to contribute 2n -1 times as much as the first cell, and the average cell n times as much as the first cell. The simple choice of $J_t$ independent of x in (0, L) makes the cells far from identical. The difference between cells n - 1 and n is that the field from the n - 1 cell is present when the nth cell radiates. The energy flux density put into the wave by the current density is the rate at which the electric field does work on the current density, $-\int E_t J_t dx$ (units of W/m$^2$). For the nth cell the contribution is

$$-\int_{x_{n-1}}^{x_n} E_t(x',\tau) J_t(x',\tau) dx' = \frac{Z_0 J_0^2}{4} (\Delta x)^2 (2n-1). \tag{26}$$

This is exactly equal to the contribution of the nth cell to the energy flux density, as required by energy conservation. Since the average value of 2n - I in the set n = 1, 2, ... , $l$ is $l$, the contribution of the average cell is $Z_0 J_0^2 (\Delta x)^2 l / 4$. The energy in our example comes from the increased source strength required to create $J_0$ in cells of higher index n.

Because the contribution of the nth cell has a multiplicative factor 2n - 1, a large n in our example might make the electric field large enough to result in a significant Schott force in the microscopic equations of motion. However, air breakdown limits E in the actual case. The analysis has shown that the factor 2n - 1 results from a self-field effect, where the cells respond to the fields created by other cells. The energy flux density contributed by each cell is accounted for by the $\vec{J} \cdot \vec{E}$ rate at which the fields do work within that cell. Coherency enters through the E by means of the 2n - 1 multiplicative factor.

The ratio of the contribution of the nth cell to the sum of the contributions by all the cells from 1 up to n - I is given by

$$\frac{\left(E_t^2 / Z_0\right)_n}{Z_0 J_0^2 (\Delta x)^2 (n-1)^2 / 4} = \frac{2n-1}{(n-1)^2}. \tag{27}$$

Thus at large n the energy flux density becomes dominated by the distant upstream generators rather than by the locally generated fields. The locally generated transverse fields never dominate the electric fields at the exit of our one-dimensional example. If the current density were specified in terms of a distribution of electrons launched into a vacuum at $\tau = 0$ with a specific $v_t$, then at later $\tau$ and for sufficiently large $n > n_s$, the fields will be high enough that self-field effects reduce $J_t$ for cells with $n > n_s$, The example illustrates that the $6.6 \times 10^6$ W/m$^2$ energy flux density generated by high-altitude EMP is a self-field effect. The energy is balanced by a corresponding loss in Compton electron kinetic energy.

Thus far our simple example has concentrated upon energy considerations. Note that the electric field is a transverse field and cannot correspond to a force along the propagation direction x. The momentum density carried by the wave must also be balanced by a reaction in our basic source that keeps the current density in the given form. The momentum density carried by the wave is

$$\vec{g} = \vec{D} \times \vec{B} = c^{-2} \vec{E} \times \vec{H} \tag{28}$$

or $g = E_t^2 / (Z_0 c^2)$ in our example. Substituting the $E_t$ in terms of x in the interval (0,L) and taking the t derivative gives

$$\partial g / \partial t = Z_0 J_0^2 x / (2c^2) \partial x / \partial t. \tag{29}$$

With $H_t = E_t / Z_0 = -J_0 x / 2$, we find $\partial g / \partial t = -J_0 B_t$, More generally, from Maxwell's equations,

$$\partial \vec{g} / \partial t = -\vec{J} \times \vec{B}. \tag{30}$$

At a given τ, the rate of gain in momentum density in the propagating wave produced by an element Δx is equal and opposite to the JXB force density acting on the current density within the element Δx experiencing the magnetic field. Our earlier estimate of $F_{rad}$ is based upon the approximately 3-μpsi peak momentum flux density being radiated. This momentum is furnished by the $\vec{J} \times \vec{B}$ force, and ultimately by a loss in electron momentum. Interaction with the self-generated magnetic field accounts for the momentum density in the propagating wave; interaction with the self-generated electric field accounts for the energy in the propagating wave.

Because the time rate of change of momentum density in the propagating wave is furnished by the $\vec{J}_t \times \vec{B}$ force density, one might think it automatic that (on an individual electron basis) the $\vec{F}_{Bgeo} = e\vec{v} \times \vec{B}_{geo}$ in the EMP case is equal to the force on the average electron caused by its interaction with the radiating fields. This is not the case.

The peak propagating B is about 2.8 $B_{geo}$. The Compton electron velocity is mainly radial rather than transverse. (In the 10-ns rise time of the Bell Laboratories waveform, a $\beta \approx 0.9$ Compton electron in $B_{geo} = 6 \times 10^{-5}$ T rotates through <15°) The $\vec{J}_t \times \vec{B}$ varies rapidly with $\tau$ while $\vec{F}_{Bgeo}$ does not. The net force exerted by the propagating fields on a Compton electron varies with the electron velocity rather than with the transverse component. The rough equivalence between the approximately $2.2 \times 10^{-15}$ N radiation-associated retarding force and the $2.9 \times 10^{-15}$ N magnitude of $\vec{F}_{Bgeo}$ is not an automatic feature of the $\vec{J}_t \times \vec{B}$ force density.

The one-dimensional example can also be developed in a spherical geometry. There the outgoing wave is given by

$$rE_t(r,\tau) = -\frac{Z_0}{2} \int_0^r r' J_t(r',\tau) dr'. \qquad (31)$$

In a region along r we consider the parameters independent of $\theta$ and $\phi$, giving a spherically diverging outgoing wave. Let

$$J_t = \begin{cases} 0, & r < r_o \\ J_0, & r_o \leq r \leq r_\ell \\ 0, & r > L + r_o \end{cases} \qquad (32)$$

Here $r_\ell = L + r_o$, $\Delta r = L/\ell$, and at the end of the nth cell, $r_n = r_o + n \Delta r$. The transverse field and the energy flux density are

$$E_t = \frac{Z_0 J_0}{4r} \begin{cases} 0, & r < r_o \\ r^2 - r_o^2, & r_o \leq r \leq r_\ell \, r \\ L^2 - r_o^2, & r \geq r_\ell \end{cases} \qquad (33)$$

$$\frac{E_t^2}{Z_0} = \frac{Z_0 J_0^2}{16} \begin{cases} 0, & r < r_o \\ (r^2 - r_o^2)/r^2, & r_o \leq r \leq r_\ell \\ (L^2 - r_o^2)^2/r^2, & r \geq r_\ell \end{cases} \tag{34}$$

In the spherical case the power per steradian is more appropriate than the W/m². Since $dA = r^2 d\Omega$, at the end of the nth cell for $n = 1,2,3,\ldots, l$,

$$\frac{dP}{d\Omega} = \frac{r^2 E_t^2}{Z_0} = \frac{Z_0 J_0^2}{16}(r_n^2 - r_o^2)^2 \tag{35}$$

The contribution from the nth cell is

$$\left(\frac{r^2 E_t^2}{Z_0}\right)_n = \frac{Z_0 J_0^2}{16}(r_n^4 - 2r_n^2 r_o^2 - r_{n-1}^4 + 2r_{n-1}^2 r_o^2) \tag{36}$$

For this nth cell the rate at which work is done by the fields (per steradian) is given by

$$-\int_{r_{n-1}}^{r_n} r'^2 E_t(r',\tau) J_t(r',\tau) dr' = \frac{Z_0 J_0^2}{8}\left[\frac{r_n^4 - r_{n-1}^4}{2} - r_o^2(r_n^2 - r_{n-1}^2)\right], \tag{37}$$

again in agreement with $\left(r^2 E_t^2 / Z_0\right)_n$, as required by energy conservation. Development of the ratio of the contribution by the nth cell to the contribution of all the cells from 1 to n - 1 is a little more tedious than for the Cartesian case. At large n, again the ratio scales as 1/n. The local transverse fields never dominate the electric fields at the exit.

Rather than using the explicit division into cells to illustrate energy and momentum conservation, we can derive these constraints from Eq. (A10). At given $\tau$ in a material of impedance $Z_0$, multiplication by the component $E_t$ parallel to J, gives

$$E_t \partial E_t / \partial x = -Z_0 J_t E_t / 2 \tag{38}$$

or

$$\int_0^x J_t E_t dx' = -E_t^2 / Z_0. \tag{39}$$

That is, the energy flux carried by the plane wave ($E_t^2 / Z_0$) at any position x is equal to the net rate at which work is done (per unit area) by the upstream transverse current density in creating the transverse electric field intensity.

The current density has convection and conduction components, $J_t = J_{ct} + \sigma E_t$ which are in opposing directions. Substitution of the components into Eq. (39) gives

$$\int_0^x J_{ct} E_t dx' + \int_0^x \sigma E_t^2 dx' = -E_t^2 / Z_0. \tag{40}$$

The rate at which the transverse convention-current density has done work on the upstream E, added to the rate at which the field loses energy to the upstream conduction electrons gives the net energy flux carried by the propagating wave at position x.

At a given retarded time, the rate of gain in momentum density carried by the wave propagating in a medium of impedance $Z_0$ is

$$\frac{\partial g}{\partial t} = \frac{\partial}{\partial t}\left(\frac{E_t^2}{Z_0 c^2}\right) = \frac{2E_t}{Z_0 c}\frac{\partial E_t}{\partial x}. \tag{41}$$

Substitution for the spatial gradient term from Eq. (38) gives $\partial g / \partial t = -J_t E_t / c$. The transverse plane wave has $E_t = cB_t$, so $\partial g / \partial t = -J_t B_t$, consistent with Eq. (30). The rate of gain in momentum density carried by the propagating wave is equal to the $J_t \times B_t$ force density acting upon the current density. Separation into convection and conduction terms gives

$$\frac{\partial g}{\partial t} = -J_{ct}B_t - \sigma E_t B_t. \tag{42}$$

Analogous to the energy flux calculation, a loss in the transverse convection-current-density term adds momentum to the propagating wave; the transverse conduction-current-density term absorbs momentum from the propagating wave.

## CONCLUSIONS

We have identified the mechanism that generates the large retarding radiation-associated force acting on the Compton electrons that produce the high-altitude EMP. The force is significant and is implicit in EMP calculations for MeV energies when current densities are modeled in a self-consistent manner. The momentum flux density in the propagating EMP results from the force due to the interaction between the self-generated magnetic field and the transverse component of current density in the gamma-absorption region of the atmosphere. As long as the transverse current density is obtained in a self-consistent manner, then the energy and momentum in the propagating wave result directly from $\vec{J}_t \cdot \vec{E}_t$ terms that drain energy and transverse momentum from the current density and from $\vec{J}_t \times \vec{B}_t$ terms that drain longitudinal momentum from the current density. When convection-electron equations of motion are solved self-consistently, the $e\vec{v} \times \vec{B}$ and the $e\vec{E}$ forces include the reaction forces acting on the transverse current density.

The relativistic form of the classical radiation reaction forces on a single particle have been examined. These are the additional forces introduced into the equations of motion just because the charged particle is radiating energy and momentum. Present high-altitude EMP codes neglect these reaction forces; hence, rigorous energy and momentum conservation are not inherent within the codes. However, radiation reaction forces as calculated for an isolated electron are negligible compared to other forces on the MeV Compton electrons. These reaction forces $F_{rr}^{\mu}$ and $S^{\mu}$ are a type of self-field effect where the electron reacts to its own radiation. Neglect of these forces does not lead to overestimates of the high-altitude EMP peak. As described in the previous paragraph, the primary self-field effect is the electron reacting to the coherent fields created by the other charged particles to provide the $\vec{J}_t \cdot \vec{E}_t$ and the $\vec{J}_t \times \vec{B}_t$ terms.

## ACKNOWLEDGMENTS

This work was benefited from helpful discussions with Fritz Rohrlich, Gary Scrivner, and Gene Salamin. It was supported by the U.S. Department of Energy and by the Electric Power Research Institute.

## APPENDIX: TRANSVERSE ELECTRIC FIELD
## FOR A ONE-DIMENSIONAL OUTGOING WAVE

In SI units, the differential forms for Maxwell's curl equations are Ampere's law,

$$\nabla \times H = J_c + \sigma E + \partial D / \partial t, \tag{A1}$$

and Faraday's law of induction,

$$\nabla \times E = -\partial B / \partial t \qquad (A2)$$

The current density is divided into the convection current $J_c$, the conduction current $\sigma E$, and the displacement current $\partial D / \partial t$. The divergence of Faraday's law gives $\partial(\nabla \cdot B)/\partial t = 0$, and so in the initial absence of magnetic monopoles, $\nabla \cdot B$ remains zero.

The equation of continuity

$$\nabla \cdot (J_c + \sigma E) + \partial \rho / \partial t = 0 \qquad (A3)$$

coupled with the divergence of Ampere's law (A1), gives $\partial(\nabla \cdot D - \rho)/\partial t = 0$. The initial condition $\nabla \cdot D = \rho$ remains true at later times.

The equations are sometimes more convenient to solve after a transformation of variables. When the EMP is produced by a propagating pulse of photons, spatial variables have steep gradients at the position of the pulse. At a given position, as the pulse passes, the fields change rapidly with both space and time. At a given retarded time the spatial variation is much less severe. This allows larger spatial steps in numerical simulations of the phenomena. In terms of retarded time $\tau = t - (\mu\varepsilon)^{1/2} r$ in a spherical coordinate system $(r, \theta, \phi)$, the various operators transform as $\nabla \to \nabla - (\mu\varepsilon)^{1/2} \hat{e}_r \partial / \partial r$ and $\partial / \partial t \to \partial / \partial \tau$. Within any region where $\varepsilon, \mu$ remain uniform, the equations become

$$\mu^{-1} \nabla \times B - (\varepsilon/\mu)^{1/2} \hat{e}_r \times \partial B / \partial t = J + \varepsilon \partial E / \partial \tau,$$

$$\nabla \times E - (\varepsilon/\mu)^{1/2} \hat{e}_r \times \partial E / \partial \tau = -\partial B / \partial \tau,$$

$$\nabla \cdot B - (\mu\varepsilon)^{1/2} \partial B_r / \partial t = 0, \qquad (A4)$$

$$\nabla \cdot E - (\mu\varepsilon)^{1/2} \partial E_r / \partial \tau = \rho / \varepsilon,$$

$$\partial \rho / \partial \tau + \nabla \cdot J - (\mu\varepsilon)^{1/2} \partial J_r / \partial \tau = 0,$$

where $J = J_c + \sigma E$. Eliminating E from the equations gives

$$(\nabla^2 B)_r + \mu (\nabla \times J)_r - 2(\mu\varepsilon)^{1/2} \frac{\partial}{\partial \tau} \left[ \frac{1}{r} \frac{\partial}{\partial r} (r B_r) \right] = 0,$$

$$(\nabla^2 B)_\theta + \mu (\nabla \times J)_\theta - (\mu\varepsilon)^{1/2} \frac{\partial}{\partial \tau} \left[ \frac{2}{r} \frac{\partial}{\partial r} (r B_\theta) - \mu J_\phi \right] = 0, \qquad (A5)$$

$$(\nabla^2 B)_\phi + \mu (\nabla \times J)_\phi - (\mu\varepsilon)^{1/2} \frac{\partial}{\partial \tau} \left[ \frac{2}{r} \frac{\partial}{\partial r} (r B_\phi) - \mu J_\theta \right] = 0$$

In terms of E, with the impedance $Z = (\mu/\varepsilon)^{1/2}$, the equations become

$$(\nabla^2 E)_r - \varepsilon^{-1}(\nabla\rho)_r - 2(\mu\varepsilon)^{1/2}\frac{\partial}{\partial\tau}\left[\frac{1}{r}\frac{\partial}{\partial r}(rE_r)\right] - Z\nabla\cdot J = 0,$$

$$(\nabla^2 E)_\theta - \varepsilon^{-1}(\nabla\rho)_\theta - (\mu\varepsilon)^{1/2}\frac{\partial}{\partial\tau}\left[\frac{2}{r}\frac{\partial}{\partial r}(rE_\theta) + ZJ_\theta\right] = 0, \quad (A6)$$

$$(\nabla^2 E)_\phi - \varepsilon^{-1}(\nabla\rho)_\phi - (\mu\varepsilon)^{1/2}\frac{\partial}{\partial\tau}\left[\frac{2}{r}\frac{\partial}{\partial r}(rE_\phi) + ZJ_\phi\right] = 0.$$

In a one-dimensional radial problem where the electric field has only a transverse component ($E_t$ equals a linear combination of $E_\phi$ and $E_\theta$), where the variables depend only upon r and $\tau$, and where the temporal variations dominate the spatial variations of $E_t$, the equation for $E_t$ becomes

$$\frac{2}{r}\frac{\partial}{\partial r}(rE_t) + ZJ_t = 0. \quad (A7)$$

At a given $\tau$, the integral form for the spherical wave is then

$$rE_t(r,\tau) = -Z\int r'J_t(r',\tau)dr'. \quad (A8)$$

The same exercise in Cartesian coordinates, where $\tau = t - x(\mu\varepsilon)^{1/2}$, gives the following equations for the components of E:

$$\nabla^2 E_x - \varepsilon^{-1}\frac{\partial\rho}{\partial x} - 2(\mu\varepsilon)^{1/2}\frac{\partial}{\partial\tau}\left[\frac{\partial E_x}{\partial x}\right] - Z\nabla\cdot J = 0,$$

$$\nabla^2 E_y - \varepsilon^{-1}\frac{\partial\rho}{\partial y} - (\mu\varepsilon)^{1/2}\frac{\partial}{\partial\tau}\left[2\frac{\partial E_y}{\partial x} + ZJ_y\right] = 0, \quad (A9)$$

$$\nabla^2 E_z - \varepsilon^{-1}\frac{\partial\rho}{\partial z} - (\mu\varepsilon)^{1/2}\frac{\partial}{\partial\tau}\left[2\frac{\partial E_z}{\partial x} + ZJ_z\right] = 0.$$

In a one-dimensional Cartesian problem where the electric field has only a transverse component $E_t = E_y$ or $E_z$, where the variables depend only upon x and $\tau$, and where temporal variations dominate the spatial variations of $E_t$, the equation for $E_t$ becomes

$$\frac{\partial E_t}{\partial x} + \frac{1}{2}ZJ_t = 0. \quad (A10)$$

At a given $\tau$, the integral form for the planar wave is

$$E_t(x,\tau) = -\frac{1}{2}Z\int J_t(x',\tau)dx'. \quad (A\,11)$$